\documentclass{llncs} 
\usepackage{url,alltt}
\usepackage{amssymb}

\usepackage{epsfig}
\usepackage{wrapfig}
\usepackage{subfigure}
\usepackage{rotating}
\usepackage{moreverb}
\usepackage{fancyvrb}

\def\bd{\begin{description}}
\def\ed{\end{description}}
\def\bc{\begin{center}}
\def\ec{\end{center}}
\def\bq{\begin{quote}}
\def\eq{\end{quote}}
\def\bi{\begin{itemize}}
\def\ei{\end{itemize}}
\def\be{\begin{enumerate}}
\def\ee{\end{enumerate}}
\def\ba{\begin{array}}
\def\ea{\end{array}}

\newcommand{\simp}{{\; \Leftrightarrow \;}}

\newcommand{\prop}{{\; \Rightarrow\; }}

\begin{document}

\title{Constraint Handling Rules - What Else?} 
\author{%
Thom Fr{\"u}hwirth
}
\institute{%
University of Ulm, Germany\\
\url{www.constraint-handling-rules.org}
}

\maketitle

\begin{abstract}

Constraint Handling Rules (CHR) is both an effective concurrent declarative constraint-based programming language and a versatile computational formalism. 
While conceptually simple, 
CHR is distinguished by a remarkable combination of desirable features:
\begin{itemize}
\item a semantic foundation in classical and linear logic, 
\item an effective and efficient sequential and parallel execution model
\item guaranteed properties like the anytime online algorithm properties 
\item powerful analysis methods for deciding essential program properties.
\end{itemize}
This overview of  
CHR research and applications is by no complete. 
It concentrates on the years since 2000. 
Up-to-date information on CHR can be found at the CHR web-site 
\url{www.constraint-handling-rules.org}, including the slides of the keynote talk associated with this article, 
online demo versions and free downloads of the language.

\end{abstract}

The final publication \cite{fruhwirth2015constraint} is available at Springer via \url{http://dx.doi.org/10.1007/978-3-319-21542-6_2}.

\section{Introduction}

Constraint Handling Rules (CHR) \cite{chr-book}
tries to bridge the gap between theory and practice, between logical specification and executable program by abstraction through constraints and the concepts of computational logic.
CHR has its {\em roots} in constraint logic programming and concurrent constraint programming, but also integrates ideas from multiset transformation and rewriting systems as well as automated reasoning and theorem proving. It seamlessly blends multi-headed rewriting and concurrent constraint logic programming into a compact user-friendly rule-based programming language.
CHR 
consists of guarded reactive rules that transform multisets of relations called constraints until no more change occurs.
By the notion of constraint, CHR does not need to distinguish between data and operations, and its rules are both descriptive and executable.

In CHR, one distinguishes two main kinds of rules: {\em Simplification
  rules}\index{rule!simplification}\index{simplification rule} replace
  constraints by simpler constraints while preserving logical equivalence, e.g.,
  \texttt{X\(\leq\)Y\(\land\)Y\(\leq\)X \(\simp\) X\(=\)Y}. {\em Propagation
  rules}\index{rule!propagation}\index{propagation rule} add new constraints
  that are logically redundant but may cause further simplification, e.g.,
  \texttt{X\(\leq\)Y\(\land\)Y\(\leq\)Z \(\prop\) X\(\leq\)Z}.
  Together with
\texttt{X\(\leq\)X \(\Leftrightarrow\) true}, these rules encode the
axioms of a partial order relation. 
The rules compute its transitive closure and
replace inequalities $\leq$ by equalities $=$ whenever possible.
For example, \texttt{A\(\leq\)B\(\land\)B\(\leq\)C\(\land\)C\(\leq\)A} becomes 
\texttt{A\(=\)B\(\land\)B\(=\)C}.
More program examples can be found in Section 2.
Semantics of CHR are discussed in Section 3.

\subsection{Powerful Program Analysis}

One advantage of a declarative programming language is the ease of {\em program analysis}. 
CHR programs have a number of desirable properties guaranteed and can be analyzed for others.  
They will be discussed in Section 4.

Since CHR (and many of its fragments) are Turing-complete, {\em termination} is undecidable, but often a ranking in the form a a well-founded termination order can be found to prove termination. From the ranking, a crude upper bound for the time complexity can automatically be derived. More precise bounds on the complexity can also be found by inspecting the rules.

{\em Confluence} of a program guarantees that any computation starting from the same initial state results in the same final state no matter which of the applicable rules are applied. There is a decidable, sufficient and necessary condition for confluence of terminating programs.

Any terminating and confluent CHR program has a consistent logical reading.
It will automatically implement a 
{\em concurrent any-time (approximation) and on-line (incremental) algorithm}, 
where constraints can arrive during the computation that can be stopped and restarted at any time.
It ensures that rules can be applied in parallel to different parts of a state
without any modification and without harming correctness.
This property is called {\em declarative concurrency} or {\em logical parallelism}.

Surprisingly, there is also a decidable, sufficient and necessary syntactic condition for {\em operational equivalence} of terminating and confluent programs (we do not know of any other programming language in practical use with this property). So one can check if two programs behave in the same way and if a program has redundant parts.

\subsection{Implementations and Applications}

CHR is often used as a language extension to other programming languages, 
its syntax can be easily adapted to that of the {\em host language}.
In the host language, CHR constraints
can be posted and inspected; in the CHR rules, host
language statements can be included.
CHR libraries are now available in almost all Prolog implementations, 
but also in Haskell, Curry, Java and C as well as in hardware. 

It has been proven that every algorithm can be implemented in CHR with {\em best known
time and space complexity}, something that is not known to be possible in other
pure declarative programming languages.  
The {\em efficiency} of the language is
empirically demonstrated by optimizing CHR compilers that compete well
with both academic and commercial rule-based systems and even classical
programming languages.
The fastest implementations of CHR, e.g. in C, allow to apply up to millions of rules per second. 

Other rule- and logic-based approaches have been successfully and rather straightforwardly embedded in CHR. For this reason, CHR is considered a candidate for a {\em lingua franca} of such approaches with the potential for cross-fertilization of research in computational systems and languages. 
Implementations and embeddings are discussed in Section 5.

CHR has been used for such {\em diverse applications} as type system design for
Haskell, time tabling, optimal sender placement, computational
linguistics, spatio-temporal reasoning, verification, semantic web
reasoning, data mining and computational linguistics. 
Successful {\em commercial application} include financial services, 
network design, 
mould design, robot vehicle control, enterprise applications
and software verification.
Applications of CHR and research using CHR are discussed in Section 6.

CHR is also available {\em online} for demos and experimentation at
\url{chrjs.net} at an introductory level 
and as WebCHR at \url{chr.informatik.uni-ulm.de/~webchr/} with more than 50 example programs. 
More than 200 academic and industrial projects worldwide use CHR, and about 200 scientific books and 2000 research papers reference it.
The CHR community and other interested researchers 
and practitioners gather at the yearly CHR workshops and the biannual CHR summer schools.

\section{A Taste of CHR Programs}

The following programs can be run with little modification in the online versions of CHR just mentioned.
Note that all programs have the anytime online algorithm properties. 
So they can be stopped at any time for intermediate results, 
constraints can be added while they already run (incrementality), 
and they can be directly executed in parallel.
These program examples are explained more in \cite{fru_welcome_lnai08} 
and discussed in detail in \cite{chr-book}. 

Some examples use a third kind of rule, a hybrid rule called simpagation rule.
It has the form 
$H_1 \backslash H_2 \Leftrightarrow C | B$.
Basically, 
if $H_1$ and $H_2$ match constraints and the guard $C$ holds, then 
the constraints matching $H_1$ are kept,
the constraints matching $H_2$ are removed and the body $C$ is added.
For logical conjunction $\land$ we will simply write a comma between constraints.\\ 

\noindent \textbf{Multiset Transformation} - \textrm{One-Rule Algorithms}
\begin{alltt}
\textrm{Compute minimum of a set of{\tt min} candidates}
min(I) \textbackslash min(J) \(\Leftrightarrow\) J>I | true.
\textrm{Compare two numbers, keep smaller one.}

\textrm{Compute greatest common divisor of a set of numbers}
gcd(I) \textbackslash gcd(J) \(\Leftrightarrow\) J>=I | gcd(J mod I). 
\textrm{Replace I and J by I and (J mod I) until all numbers are the same.}

\textrm{Compute primes, given{\tt prime(2),...,prime(MaxN)}}
prime(I) \textbackslash prime(J) \(\Leftrightarrow\) J mod I = 0 | true.
\textrm{Keep removing multiples until only primes are left.}

\textrm{Sort array with elements{\tt a(Index,Value)}}
a(I,V), a(J,W) \(\Leftrightarrow\) I>J, V<W | a(I,W), a(J,V).
\textrm{Keep swapping numbers that are out of order until sorted.}

\textrm{Merge Sort, given values as{\tt next(start,Value)}}
next(A,B) \textbackslash next(A,C) \(\Leftrightarrow\) A<B,B<C | next(B,C).
\textrm{Turn common successors into direct successors until sorted chain results.}

\textrm{Newton's Method for Square Root Approximation for{\tt N>1}}
eps(E) \textbackslash sqrt(X,R) \(\Leftrightarrow\) R*R/X-1>E | sqrt(X,(R+X/R)/2).
\textrm{Start with{\tt sqrt(N,N)}.{\tt E} is the required precision factor.}
\end{alltt}

\noindent {\bf Fibonacci Variations}
- {\tt M} is the {\tt N}th Fibonacci number
\begin{alltt}
\textrm{Top-down Evaluation}
fib(0,M) \(\Leftrightarrow\) M=1.              
fib(1,M) \(\Leftrightarrow\) M=1.
fib(N,M) \(\Leftrightarrow\) N>=2 | fib(N-1,M1), fib(N-2,M2), M=M1+M2.
\textrm{Matching is used on left hand sides of rules.}

\textrm{Top-down Evaluation with Memorization (in first rule)}
fib(N,M1) \textbackslash fib(N,M2) \(\Leftrightarrow\) M1=M2.
fib(0,M) \(\Rightarrow\) M=1.              
fib(1,M) \(\Rightarrow\) M=1.
fib(N,M) \(\Rightarrow\) N>=2 | fib(N-1,M1), fib(N-2,M2), M=M1+M2.
\textrm{Turned simplification into propagation rules.}

\textrm{Bottom-up Evaluation without Termination}
fibstart \(\Leftrightarrow\) fib(0,1), fib(1,1).
fib(N1,M1), fib(N2,M2) \(\Rightarrow\) N2=N1+1 | fib(N2+1,M1+M2).
\textrm{Basically, original simplification rules have been reversed.}

\textrm{Bottom-up Evaluation with Termination at{\tt Max}}
fib(Max) \(\Rightarrow\) fib(0,1), fib(1,1).
fib(Max), fib(N1,M1), fib(N2,M2) \(\Rightarrow\) Max>N1, N1=N2+1 | 
                              fib(N2+1,M1+M2).
\textrm{The auxiliary constraint{\tt fib(Max)} is added. Computation stops when{\tt Max=N1}.}
\end{alltt}

\noindent {\bf All-Pair Shortest Paths}
\begin{alltt}
\textrm{The distance from \texttt{X} to \texttt{Y} is \texttt{D}}
path(X,Y,D1) \textbackslash path(X,Y,D2) \(\Leftrightarrow\) D1=<D2 | true.
arc(X,Y,D) \(\Rightarrow\) path(X,Y,D).
arc(X,Y,D), path(Y,Z,Dn) \(\Rightarrow\) path(X,Z,D+Dn).
\textrm{Compute all paths with propagation rules, keep smaller ones.}
\end{alltt}

\noindent {\bf Dynamic Programming}
 - Bottom-up Parsing with CYK Algorithm
\begin{alltt}
\textrm{{Grammar rules} are in Chomsky normal form{\tt A->T} or{\tt A->B*C}.}
\textrm{A sequence of terminal symbols is encoded as a chain of arcs.}
parse(X,Y,A) \textbackslash parse(X,Y,A) \(\Leftrightarrow\) true.
terminal @ A->T, arc(X,Y,T) \(\Rightarrow\) parse(X,Y,A).
non-term @ A->B*C, parse(X,Y,B), parse(Y,Z,C) \(\Rightarrow\) parse(X,Z,A).
\textrm{Note the similarity with All-Pair Shortest Paths.}
\end{alltt}

\noindent {\bf Boolean Conjunction as Constraint}
 \begin{alltt}
\textrm{The result of \texttt{X}\(\land\)\texttt{Y} is \texttt{Z}}
and(X,Y,Z) \(\Leftrightarrow\) X=0 | Z=0.     and(X,Y,Z) \(\Leftrightarrow\) Y=0 | Z=0.
and(X,Y,Z) \(\Leftrightarrow\) X=1 | Z=Y.     and(X,Y,Z) \(\Leftrightarrow\) Y=1 | Z=X.
and(X,Y,Z) \(\Leftrightarrow\) X=Y | Y=Z.     and(X,Y,Z) \(\Leftrightarrow\) Z=1 | X=1,Y=1.
\textrm{Also computes with unknown input values and backwards. Such rules can be automatically generated from specifications \cite{generate}.}
 \end{alltt}

\section{CHR Semantics}

In this section we give an overview of the main semantics for CHR.
More detailed overviews can be found in \cite{chr-thesis-book,betz2014unified}.
As a declarative programming language and formalism, CHR features both operational semantics that describe the execution of a program and declarative semantics that interpret a program as a logical theory. These semantics exist at various levels of refinement. They are related by soundness and completeness results, showing their correspondence.

\subsection{CHR Rules and their Declarative Semantics}

To simplify the presentation, we use a generic notation for all three kinds of CHR rules.
{\em Built-in constraints} are host language statements that can be used as tests in the guard or auxiliary computations in the body of a rule.
A {\em generalized simpagation rule} is of the form
\[H_1 \backslash H_2 \Leftrightarrow C | B\]
where in the rule head (left-hand-side), $H_1$ and $H_2$ are conjunctions of user-defined constraints,
the optional guard $C$ is a conjunction of built-in constraints from the host language
and the body (right-hand-side) $B$ is a conjunction of arbitrary constraints.
If $H_1$ and $H_2$ are non-empty, the rule corresponds to a simpagation rule.
If $H_1$ is empty, the rule corresponds to a simplification rule,
if $H_2$ is empty, the rule corresponds to a propagation rule.

The declarative semantics is based on first-order predicate logic, 
where constraints are viewed as predicates and rules as logical implications and equivalences. 
A generalized simpagation rule 
basically corresponds to a logical equivalence
\[H_1 \land H_2 \land C \leftrightarrow H_1 \land C \land B.\]
An interesting refinement is the {\em linear-logic semantics} \cite{betz2014unified,betz2013linear}. 
It is closer to the operational semantics in that it captures the meaning of constraints as resources, 
where multiplicities matter.

\subsection{Operational Semantics for CHR}

The execution of CHR can be described by structural operational semantics, 
which are given as state transition systems. 
Basically, states are conjunctions of constraints.
These semantics exist in various formulations and at various levels of refinement,
going from the abstract (analytical) to the concrete (pragmatic): 
\begin{itemize}
\item The {\em very abstract semantics} \cite{chr-book} is close to modus ponens of predicate logic. 
\item The {\em abstract (or theoretical) semantics} \cite{Abdennadher+99} is often used as a basis 
for program analysis. 
\item The {\em refined semantics} \cite{duck_stuck_garc_holz_refined_op_sem_iclp04} describes the behavior of CHR implementations.
\end{itemize}
Several alternative operational semantics for CHR have also been proposed, among them
\cite{gabbr_meo_compos_semantics_tocl09,haemmerle2011coinductive,sarnastarosta_ramakrishnan_chrd_padl07,betz_raiser_fru_execution_model_iclp10}.

The essential aspect of the operational semantics is the application of a rule:
Take a generalized simpagation rule from the program.
If there are constraints in the current state that match the head of the rule and if the guard holds under this matching, then the constraints matching second part of the head 
$H_2$ (if any) are removed and the guard and body of the rule are added to the state.

There are alternative formulations for the above semantics. 
Chapter 8 in the book \cite{chr-thesis-book}
and 
\cite{raiser_betz_fru_equivalence_revisited_chr09,betz2014unified} develop an axiomatic notion of state equivalence.
The equivalence relation $\equiv$ on states treats built-in constraints semantically and user-defined constraints syntactically. Basically, two states are equivalent if they are logically equivalent while taking into account that - forming multisets - multiplicities of user-defined constraints matter.
For example, $X{=<}Y \land Y{=<}X \land c(X,Y) \ \equiv \ X{=}Y \land c(X,X)$ which is different to $X{=}Y \land c(X,X) \land c(X,X)$.

Using state equivalence, the presentation of the abstract semantics can be simplified. 
It basically boils down to

\begin{center}
$\underline{S \equiv (H_1 \land H_2 \land C \land G) \ \ \ \ (H_1 \backslash H_2 \Leftrightarrow C | B)  \ \ \ \ (H_1 \land C \land B \land G) \equiv T}$\\
$S \longmapsto T$
\end{center}

where all upper-case letters stand for conjunctions of constraints.
$G$ is called the context of the rule application, $G$ is not affected by it.
Note that the transition $S \longmapsto T$ is only allowed if the built-in constraints in state $S$ are consistent and 
if the rule has not been applied before to the same constraints under the same matching.

\subsection{Operational Semantics for Parallel CHR}

One of the main features of CHR is its inherent \index{concurrency}{\em concurrency}.
Intuitively, in a parallel execution of CHR we can apply rules simultaneously to different parts of a state. But we can do more than that: We can also apply rules to overlapping parts of a state as long as the overlap is only removed by at most one rule.
In Chapter 4 of \cite{chr-book}, this parallelism in CHR is defined by an interleaving semantics as

\begin{center}
$\underline{A \land G \longmapsto C \land G \ \ \ \ \ \ B \land G \longmapsto D \land G}$\\
$A \land B \land G \longmapsto C \land D \land G$
\end{center}

This inference rule is justified by the monotonicity property of CHR (explained below).
If a program executed under the refined semantics makes use of the order of constraints in a state and the order of rules in a program, this kind of automatic parallelization may not work.
Such programs are not {\em confluent}. 
On the other hand, confluent programs can be executed in parallel without modification.
As we will see, we can check CHR programs for confluence, and we can even semi-automatically complete them to make them confluent. 
Thus, using completion, we can turn non-confluent programs into parallel programs.
This method has been applied to the classical Union-Find algorithm which is very hard to parallelize
\cite{fru_parallel_union_find_iclp05} 
(with \cite{triossi2012compiling} showing the effectiveness of the resulting program) 
and to the
Preflow-Push algorithm \cite{meister_preflow_push_wlp06}.
Alternative and more refined semantics for parallel CHR are e.g.
\cite{lamsulz08,schr_sulz_transactions_iclp08,lam_sulzmann_conc_goal_based_tplp09,raiser_fru_parallel_wlp10,chr-thesis-book}.

\section{Properties of CHR and Their Analysis}

We first introduce three essential types of monotonicity and the anytime online algorithm properties that all come for free in CHR. 
We then discuss the analysis of termination and time complexity as well as of confluence, completion and operational equivalence
of CHR programs.

\subsection{CHR Monotonicity Properties}

In the abstract operational semantics we can observe three essential types of monotonicity.

First, adding rules to a program cannot inhibit the applicability of any rules that were applicable. This aids incremental program development and rapid prototyping. Already a program with a few first rules is executable, and we can add rules to cover more and more cases, enabling more and more desired computations. The confluence test (see next Section) can be used to discover situations where old and new rules lead to different results.

Second, built-in constraints (that occur in the guard and body of a rule) can only be added to a state, they are never removed. Hence they accumulate monotonically.
On the other hand, user-defined constraints are non-monotonic in that they can be added and removed from a state.
This means that an applicable rule will remain applicable as long as the user-defined constraints it matches are present in the state and as long as the state is consistent.

Third, during a rule application, the context $G$ stays unchanged. We can actually change it without influencing the rule application itself.
So if a rule is applicable in a state, it is also applicable in any larger state where constraints have been added (as long as the state is consistent) \cite{Abdennadher+99}.
This is an important {\em modularity property} of CHR, it is usually called CHR's {\em monotonicity property}. 
Clearly such context-independence does not hold in traditional programming languages, where the context may update as well, resulting in write conflicts.

On the other hand, if we have an empty context $G$, we get the {\em minimal transition} for to the given rule:
\[(H_1 \land H_2 \land C) \longmapsto (H_1 \land C \land B).\]
The state $(H_1 \land H_2 \land C)$ is called {\em minimal state} of the rule. 
Removing any constraint from it would make its rule inapplicable. 
Adding constraints to it cannot inhibit the applicability due to monotonicity.
Since minimal states and transitions capture the essence of a rule application, they will come handy later when analyzing CHR programs for confluence and operational equivalence.

\subsection{Anytime Online Algorithm Properties}

Any algorithm expressed properly as a CHR program will enjoy several important properties:
It will be an anytime algorithm and it will be an online algorithm and it can be run in parallel without modification.

The {\em anytime (approximation) algorithm property} means that we can interrupt
the execution of a program at any time, observe the current state as 
an approximation to the result and restart from that
intermediate result. This is obvious from the operational semantics and the notion of states  
and transitions used there.

The {\em online (incremental) algorithm property} means that we can add
additional constraints while the program is running without the need to
recompute from scratch. 
This is an immediate consequence of the monotonicity property of CHR.
The program will behave as if the newly added constraints were present from the beginning but had been ignored so far. Therefore only a minimal amount of computation is performed to accommodate the new constraint.
Incrementality is useful for interactive, reactive and control systems, 
in particular for agent and constraint programming.

In the refined semantics, the order of constraints in a state and the order of rules in a program can be made to matter, and this may weaken the above properties.

\subsection{Termination and Time Complexity Analysis}

One way to show termination is to prove that in
each rule, if the guard holds, the rule head is strictly larger than
the rule body using some well-founded termination order called a ranking. 
For CHR programs that mainly use simplification rules, simple
rankings are often sufficient to prove termination~\cite{Fruehwirth00,FruKR02}. 
More sophisticated methods are needed in the presence of propagation rules \cite{pilozzi_auto_term_proofs_iclp09,pilozzi_improved_termination_lopstr11,thom2015ppdp}.
An approximation of CHR programs by constraint logic programs (CLP) 
has also been used to analyse the termination behavior of CHR \cite{haemmerle2011clp}. 

The \index{run-time}run-time of a CHR program not only depends on the number of
rule applications (derivation lengths), but also on the
number of \index{rule application attempt}\index{rule try}rule application
{attempts}.
The meta-complexity theorem in \cite{Fruehwirth2002} basically states that the complexity is bounded by the derivation length taken
to the power of the number of heads in a rule.
This only gives crude upper-bounds.

Actual CHR systems achieve much better complexity results since they implement the refined
semantics and feature compiler optimizations such as
indexing.
For CHR with and without priorities, there is a more realistic sophisticated 
meta-complexity result derived from the Logical Algorithms (LA) formalism \cite{dekoninck_la_meets_chr_tplp09}.

\subsection{Confluence and Completion} 

Confluence means that it does not matter for the result which of the applicable
rules are applied in which order in a computation. 
The resulting states will always be equivalent to each other.
For terminating CHR programs, there is a decidable, sufficient and necessary
condition for confluence~\cite{Abdennadher+99}. 
These papers also have shown the many benefits of confluent programs:
\begin{itemize}

\item Confluent programs are always implement anytime online algorithms.

\item Confluent programs can be run in parallel without modification.

\item Confluence implies consistency of the logical
reading of the program. 

\item Confluence improves the soundness and completeness results between the operational and
declarative semantics. 
These theorems are stronger than those for other (concurrent) constraint programming languages.

\item The least models of confluent CHR programs and its CLP approximation coincide \cite{haemmerle2011clp}.
\end{itemize}

The idea of the confluence test is to construct a finite number of so-called {\em critical states} by overlapping minimal states of rules in the program. An overlap equates some user-defined constraints and removes the resulting duplicate occurrences. If these constraints are to be removed by more than one rule, we have generated a conflict. One now checks if these conflicting rule applications on its own can be continued with computations that lead to equivalent states. If this holds for all critical states in the program, we have proven confluence.

In practice, this notion of confluence can be too strict.
In~\cite{duck_stuck_sulz_observable_confluence_iclp07} the notion of \index{observable
confluence}{\em observable confluence} is
introduced, where the states considered must
satisfy a user-defined \index{invariant}{invariant}. 
Other related notions of confluence are considered in \cite{HaemmerlE:2012:DCC:2431176.2431194,christiansenconfluence}.
Confluence for non-terminating programs is in general undecidable, it is discussed in \cite{raiser_tacchella_confluence_non_terminating_chr07}.

{\em Completion} is the process of adding rules to a non-confluent program
until it becomes confluent~\cite{abd_fru_completion_cp98}. 
These rules are generated between the successor states of critical states.
In contrast to completion for term rewriting, in CHR we generally need more than
one rule to make a critical pair joinable: a simplification rule and a
propagation rule. Unfortunately, completion may not terminate.
Completion can be also used for program specialisation 
\cite{abd_fru_integration_lopstr03,abd_fru_completion_cp98}.

\subsection{Operational equivalence}

Operational equivalence means that given two programs, for any given state, its computations in both
programs lead to the same final state. 
There is a decidable, sufficient and necessary condition for
operational equivalence of terminating and confluent 
CHR programs~\cite{abd_fru_equivalence_cp99}. We do not know of any
other programming language in practical use that admits such a test.

The test is straightforward: 
The minimal states of the rules in both programs are each executed in both programs, and for each minimal
state, the computations must reach equivalent states in both programs.  
This test can also be used to discover redundant rules in a program.

\section{CHR Implementations and Embeddings in CHR}

We discuss efficient implementations, variants and extensions of CHR and embeddings of other rule- and graph-based approaches in CHR.

\subsection{CHR Implementations and Their Efficiency}

The first wide-spread implementations of CHR were based on \cite{holz_fru_prolog_chr_compiler_aai00}.
Most available CHR implementations today - be it in Prolog, Java or C - 
are based on the expertise of the CHR team at Katholieke Universiteit Leuven
\cite{wuille_schr_demoen_cchr_chr07,vanweert_wuille_et_al_chr_imperative_lnai08,vanweert_jchr_compilation_techrep08}.

State-of-the-art CHR libraries with mode and type declarations in Prolog and C allow to implement any algorithm in a natural and high-level way, with time and space consumption that is typically 
within an order of magnitude from the best-known implementations in any other language \cite{sney_schr_demoen_chr_complexity_toplas09,vanweert_lazy_evaluation_tkde10}.
Indeed, \cite{sney_schr_demoen_chr_complexity_toplas09} has proven that {\em every algorithm can be implemented in CHR with the best known time and space complexity}. 
This has been exemplified by providing elegant implementations with optimal time-complexity of the classical union-find algorithm \cite{schr_fru_opt_union_find_tplp06} and Fibonacci heaps \cite{sney_schr_demoen_dijkstra_chr_wlp06}. 
CHR is the only known declarative language where this results holds, it is unlikely to hold for other declarative languages like Prolog or Haskell \cite{sney_schr_demoen_chr_complexity_toplas09}.
Actually, CHR cannot be embedded in pure Prolog \cite{GabbrielliMM13}. 
The fastest CHR implementations in CCHR \cite{wuille_schr_demoen_cchr_chr07} and hProlog allow to up to
apply millions of rules per second.

One reason for the effectiveness of CHR is that it uses a compiler and run-time system that is a significant advancement over existing algorithms (such as RETE, TREAT, LEAPS) for executing rule-based languages as has been impressingly demonstrated in \cite{vanweert_lazy_evaluation_tkde10}.
In addition to a superior rule-application mechanism, CHR compilers use sophisticated optimizations 
(besides indexing on constraint arguments taking into account mode and type information), 
such as memory reuse, late storage, guard optimization and join ordering optimization 
\cite{holz_garc_stuck_duck_opt_comp_chr_hal_tplp05,vanweert_lazy_evaluation_tkde10,chr-thesis-book}.

CLIPS (in C) and JESS (in Java)) are considered by many to be the most efficient rule-based systems available.
The benchmarks of \cite{vanweert_lazy_evaluation_tkde10} show that his novel
Java implementation of CHR as well as CHR in C (CCHR) \cite{wuille_schr_demoen_cchr_chr07} 
are faster than CLIPS and JESS, sometimes by several orders of magnitude. 
In benchmarks of \cite{sney_schr_demoen_chr_complexity_toplas09}, 
CHR with mode declarations achieves the optimal time and space complexity (as do imperative languages).
Prolog and strict Haskell 
have a time complexity which is a polylogarithmic factor from optimal, and their space complexity is not optimal. 
Lazy Haskell quickly gets into memory problems.

As for concurrency, prototype {\em parallel CHR} implementations exist in software using Haskell \cite{lamsulz08} 
and in hardware using Nvidia CUDA by transforming a subset of CHR to C++ \cite{zaki_parallel_gpu_chr12}
and using FPGA's \cite{triossi2012compiling}. 
These papers feature experiments that show a potential for optimal linear speedup by parallelization of CHR programs 
(and super-linear speed-up e.g. in the case of the greatest-common-divisor program).

\subsection{CHR Language Variants and Extensions}

We start with a remark on fragments of CHR, indicating the adequacy of the overall language.
We then discuss language extensions for CHR, 
program transformation 
and new programming languages based on CHR.

While there are many Turing-complete language subsets of CHR \cite{sneyers_subclass_iclp08,gabbrielli_mauro_meo_sneyers_dedidability_iclp10,mauro2014constraint} 
(a single multi-headed simplification rule suffices), it has also been shown in \cite{giustu_gabbri_meo_multiple_heads_sofsem09,GabbrielliMM13} that each of the following features of CHR can be considered essential, since they increase the expressive power of CHR: constraints with arguments, built-in constraints, function symbols to build complex terms, multi-headed rules, introduction of new variables in the body of a rule.

Since CHR libraries in Prolog naturally allow to use backtracking {\em search} by Prolog's disjunction,
most operational semantics can be extended to the resulting language CHR$^\vee$ \cite{AbdennadherSchuetz98}.
In \cite{dekoninck_schr_demoen_search_chr06} the authors extend the refined operational semantics of CHR
to support the implementation of different search strategies. 

In {\em adaptive} CHR, constraints can be declaratively removed together with the 
consequences they produced by getting involved in rule applications. 
This means that any properly written algorithm becomes adaptive.
An adaptive semantics is defined in \cite{wolf_robin_vitorino_chrv_semantics_lnai08}.
Adaptive CHR is used for realizing intelligent search strategies in \cite{wolf2005intelligent,wolf_robin_vitorino_chrv_semantics_lnai08}.

In \cite{dekoninck_schr_demoen_chrrp_techrep07} the authors extend CHR
 with user-defined rule {\em priorities} that can be static or dynamic. 
	This language extension reduces the level of non-determinism that is inherent 
	to the abstract operational semantics of CHR, and gives a more high-level form 
	of execution control compared to the refined operational semantics. 
Priorities make CHR more expressive.

Other notable {\em extensions} of CHR include 
non-monotonic negation-as-absence \cite{vanweert_sney_schr_demoen_negation_chr06},
aggregates such as sum, count, findall, and min \cite{sney_vanweert_demoen_aggregates_chr07},
rules with probabilities \cite{fru02,christiansen2008implementing,sneyers2010chr},
Except for search, all above CHR extensions have been implemented by simple effective {\em source-to-source program transformation} in CHR itself, 
also see Chapter 6 in \cite{chr-book} and the
online transformation tool at \url{http://pmx.informatik.uni-ulm.de/chr/stssemantics/}.
Program transformation in itself has been studied in 
\cite{fru_holz_source2source_agp03,abd_sts_lopstr13}.
Partial evaluation is covered by
\cite{fru_specialization_lopstr04}, discussing specialisation of CHR rules, 
and by \cite{gabbrielli_unfolding_tplp13}, which is concerned with unfolding of CHR rules.
Confluence completion can be used to great effect for program specialisation \cite{abd_fru_integration_lopstr03,abd_fru_completion_cp98}.

Notable {\em new programming languages} that are based on CHR are:
\begin{itemize}

\item HYPROLOG \cite{christiansen_dahl_hyprolog_iclp05} as an extension of Prolog with assumptions and abduction.

\item DatalogLB adds features of CHR to Datalog \cite{green2012logicblox}.

\item CHRISM is CHR with probabilistic reasoning and statistical learning \cite{sneyers2010chr}.

\item CADMIUM is an implementation of ACD Term Rewriting, 
a generalization of CHR and Term Rewriting (TRS) \cite{duck_cadmium_iclp08}.

\item SMCHR is an implementation of Satisfiability Modulo Theories 
(SMT) \cite{DBLP:journals/corr/abs-1210-5307}, where the theory part can be implemented in CHR. 

\item Linear Meld (LM) is a linear logic language closely related to CHR \cite{lm-rocha2015}.

\item CoMingle is CHR for distributed logic programming (on Android) \cite{lam2015}.
\end{itemize}

\subsection{Embedding Other Formalisms and Languages in CHR}

The expressiveness, effectiveness and efficiency of CHR enables the embedding of the characteristic features of other rule-based and graph-based formalisms, systems and languages in CHR by simple {\em source-to-source transformations}:
\begin{itemize}

\item Prolog and Constraint Logic Programming (CLP) programs are translated into CHR$^\lor$
in~\cite{AbdennadherSchuetz98} using Clark's completion.

\item Logical Algorithms (LA) are mapped into CHR with and without rule priorities 
in \cite{LACHR}. This are the only known implementations of LA. They achieve the tight time complexity
required for the LA meta-complexity theorem to hold.

\item Term Rewriting Systems (TRS) are translated to rules with equational constraints in CHR in \cite{raiser_fru_towardstermrewriting_chr08}.

\item Graph Transformation Systems (GTS) are encoded in CHR in \cite{raiser08}.
Soundness and completeness of the encoding is proven. GTS joinability of critical pairs can be
mapped onto joinability of specific critical pairs in CHR.

\item Petri Nets are translated to CHR in~\cite{betzPN}.  It is proven
that there is a one-to-one correspondence between Colored Petri Nets and 
positive ground range-restricted CHR simplification rules over finite domains.

\end{itemize}
Chapter 6 and 9.3 of \cite{chr-book} 
and the CHR web-page 
also describe these embeddings:
\begin{itemize}

\item Production Rules and Business Rules,

\item Event-Condition-Action (ECA) Rules,  

\item Functional Programming, 

\item General Abstract Model for Multiset Manipulation (GAMMA),

\item Deductive databases languages like DATALOG,

\item Description logic (DL) with OWL- and SWRL-style rules,

\item Concurrent Constraint Programming (CC) language framework.

\end{itemize}
The online tool \url{http://pmx.informatik.uni-ulm.de/chr/translator} supports the basic translation for some of these embeddings:
term rewriting systems,
functional programming,
multiset transformation,
production rules with negation-as-absence.

The embeddings are quite useful for comparing and for cross-fertilization between different approaches.
For example, in the CHR embedding, the close relationship between colored Petri Nets and the GAMMA chemical abstract machine (CHAM) can be immediately seen. 
On the other hand, it seems difficult to come up with an embedding of full CHR in one of the afore-mentioned formalisms. 
Basically, other approaches either lack the notion of constraints and logical variables
or they lack multi-headed rules and propagation rules.
Given these embeddings and its power in general, CHR can be considered a candidate for a {\em lingua franca} for computational systems with the potential for cross-fertilization of research.

\section{CHR in Research and Applications}

Typical research applications of CHR can be found in areas of computational linguistics, 
constraint solving, cognitive systems, spatio-temporal
reasoning, agent-based systems, bio-informatics, semantic web, 
type systems, verification and testing and many more. 

Commercial applications include 
financial services in stockbroking (SecuritEase, New Zealand), 
vehicle control by robotic brains (Cognitive Systems, Spain), 
injection mould design (Cornerstone Intelligent Software Corp, Canada),
optical network design (Mitre, USA), 
enterprise applications (LogicBlox, USA),
and software verification (BSSE, Germany). 
See Section 7 in \cite{chr_survey_tplp10} for details.

\subsection{Language Design and Algorithm Design}

One of the most successful research applications of CHR is in the design, prototyping and analysis
of advanced type systems for the functional programming language
\index{Haskell}Haskell~\cite{sulz_duck_peyton_stuck_func_dep_via_chr_fp07,StuSu05,DBLP:journals/tplp/DuckHS14}. 
Type reconstruction with CHR is performed for functional and logic programs in~\cite{SchriBru06}. 
A flow-based approach for a variant of parametric polymorphism in Java is based on CHR
in~\cite{ChiCra+06}.

		The union-find algorithm can be seen as solving simple equations between variables or constants. 
		By choosing the appropriate equational relations, one can derive fast incremental algorithms 
		for solving certain propositional logic (SAT) problems and polynomial equations in two variables \cite{fru_general_union_find_csclp07}.
Almost-linear tree equation solving algorithms are reconstructed with CHR in \cite{meister2007reconstructing}.
Parallelizing classical algorithms is discussed for Union-Find using confluence analysis \cite{fru_parallel_union_find_iclp05} 
and for Preflow-Push \cite{meister_preflow_push_wlp06}.

\subsection{Software Verification and Testing}

The authors of \cite{gerlich_atdg_chr10,gerlich2014automatic}
present a new method for automatic test data generation
		(ATDG) applying to semantically annotated control-flow graphs (CFGs),
		covering both ATDG based on source code and assembly or virtual machine
		code. The method supports a generic set of test coverage criteria, including
		all structural coverage criteria currently in use in industrial software test for
		safety critical software.
The work \cite{aichernig_mutation_testing_2010}
gives test cases a denotational semantics by viewing them as specification predicates. 
The authors develop a testing theory and implementation for fault-based mutation testing.  

Other applications of CHR in testing include 
\cite{pretschner_et_al_model_based_testing_jss04,gouraud_gotlieb_javacard_padl06,ss_stirewalt_dillon_checking_deadlock_ijseke07,degrave_schr_vanhoof_test_inputs_merc_lopstr08}.
An an effective methodology for verifying properties of imperative programs is their transformation to constraint-based programs \cite{DBLP:conf/cp/DuckJK13,albert_clp_heap_iclp13,pettorossi2014program}. 
Somewhat related is lightweight string reasoning for OCL \cite{buttner2012lightweight}.

\subsection{Constraints Solving and Reasoning}

CHR was originally designed to write or even automatically generate constraint solvers
\cite{abd_rigo_automatic_gen_of_solvers_tocl04,generate,sobhi_abd_betz_intentionally_def_constraints_lnai08,raiser_globalconstraintautomata_cp08}.
Solvers written in CHR and applications of CHR in constraint reasoning can be found in \cite{fru_abd_essentials_of_cp_book03} and further references in 
\cite{fru_chr_overview_jlp98,chr_survey_tplp10,chr-thesis-book}.
For example, CHR-based spatio-temporal reasoning 
is applied to robot path planning in \cite{EsTo98,martinez2012general}.
In the soft constraints framework \cite{bistarelli_fru_marte_rossi_soft_constraint_propagation_CI04,DBLP:journals/cma/BistarelliMS12,bistarelli2010formal}, constraints and partial assignments are given preference or importance levels,
and constraints are combined according to combinators which express the desired optimization criteria. 

The goal of argumentation-based {\em legal reasoning} \cite{sneyers2013probabilistic} is to
determine the chance of winning a court case, given the probabilities of the judge accepting
certain claimed facts and legal rules.
In {\em computer linguistics},
CHR Grammars (CHRG) \cite{christiansen2005chr}
execute as robust bottom-up parsers with an inherent treatment of ambiguity.
{\em Computational Cognitive Modeling} is a research field at the interface of computer
science and psychology. It enables researchers to build detailed cognitive
models using a cognitive architecture. 
A popular cognitive architecture, ACT-R, has been implemented in CHR and 
given a proper formal semantics for the first time \cite{gall2015lopstr,gall2015ppdp}.

\subsection{Multi-Agent Systems and Abduction}

The agent-based system FLUX is implemented in CHR~\cite{Flux05,thielscher2006reasoning}.  Its
application FLUXPLAYER~\cite{Fluxpl07} won the General Game Playing competition
at the AAAI conference in 2006.  
SCIFF is a framework to specify and verify interaction in open agent societies \cite{alberti2008verifiable,alberti2013chr}. 
The SCIFF language is equipped with a semantics based on abductive logic programming.
Other applications in multi-agent systems and
abductive reasoning are for example~\cite{SeBa+02,AlDa+04,montali2010abductive,gavanelli_alberti_lamma_abduction_optimization_ecai08}.
HYPROLOG \cite{christiansen_dahl_hyprolog_iclp05}
extends Prolog with CHR rules for assumptions, abduction and
integrity constraints.
Probabilistic Abductive Logic Programs (PALPs) are introduced and and implemented
in CHR for solving abductive problems providing minimal
explanations together with their probabilities \cite{christiansen2008implementing}.

\subsection{Semantic Web}

In Chapter 9.3. of \cite{chr-book} a straightforward and effective implementation of description logic with OWL-
and SWRL-style rules in CHR is given. 
For the Semantic Web\index{Semantic Web}, the integration and combination of
data from different information sources is an important issue that can be
handled with CHR~\cite{BaDo+04,zhu_madnick_siegel_IJEB08}. 
In \cite{baryannis2014fluent} a composition and verification framework for Semantic Web Services specified using WSSL is proposed, a novel service specification language based on the fluent calculus, that addresses issues related to the frame, ramification and qualification problems.
An earlier paper on web service composition using fluent calculus is \cite{salomie2010web}.
The paper \cite{6274053} proposes a service modeling approach consisting of service contracts and a process model. Service contracts are used as service advertisement and service request in this approach. 
The Cuypers Multimedia Transformation
Engine~\cite{GevOs+01} supports the automatic generation of Web-based
presentations adapted to the user's needs.

\subsection{The Diversity of CHR Applications}

Scheduling and timetabling are popular constraint-based applications, and this
also holds for CHR implementation of course scheduling and room planning for the University of Munich
\cite{AbMa00,abd_saft_will_classroom_assignment_paclp00}, 
which has become an often-cited standard work in the area.

The tool Popular \cite{fru_bri_base_wireless_ieee00}  
uses a path-loss model to describe radio-wave transmission and constraint-based programming to optimize the placement of base stations (transmitters) for local wireless communication at company sites.

The Munich Rent Advisor \cite{fru_abd_munich_rent_advisor_tplp01} allows the calculation of the
estimated fair rent for a flat based on statistical data using an online form. Simply by translating the calculation scheme into CHR-based arithmetic interval constraints, the functionality is significantly extended: The user need not answer all questions, and so an interval range for the possible rent is returned.

The papers \cite{apopcaleaps,geiselhart_et_al_mtseq_chr10} present a new system for automatic music generation, in which music is modeled using very high
level probabilistic rules in CHRISM \cite{sneyers2010chr}. 
The probabilistic parameters can be learned from examples, resulting in a system for personalized music generation.

The authors of \cite{langbein_routing_robsail11}
present an algorithm for long-term routing of autonomous sailboats. It is based on the A*-algorithm and incorporates changing weather conditions by dynamically adapting the underlying routing graph. The software also takes individual parameters of the sailboat into account, and proved to be faster than commercial systems. The system was successfully put to test during an attempt to break the world record in long-distance robot sailing with the ASV RoBoat of INNOC (Vienna).

\bibliographystyle{abbrv} 
\bibliography{CHR2015,chr-book,devils,tfall2005,biblio}

\begin{thebibliography}{100}

\bibitem{abd_sts_lopstr13}
S.~Abdennadher, G.~Fakhry, and N.~Sharaf.
\newblock Towards the implementation of source-to-source transformation tool
  for {CHR} operational semantics.
\newblock In G.~Gupta, editor, {\em LOPSTR '13, Pre-proceedings}, 2013.

\bibitem{abd_fru_completion_cp98}
S.~Abdennadher and T.~Fr\"{u}hwirth.
\newblock On completion of {C}onstraint {H}andling {R}ules.
\newblock In M.~J. Maher and J.-F. Puget, editors, {\em CP '98}, volume 1520 of
  {\em LNCS}, pages 25--39. Springer, Oct. 1998.

\bibitem{abd_fru_equivalence_cp99}
S.~Abdennadher and T.~Fr{\"u}hwirth.
\newblock Operational equivalence of {CHR} programs and constraints.
\newblock In J.~Jaffar, editor, {\em CP '99}, volume 1713 of {\em LNCS}, pages
  43--57. Springer, Oct. 1999.

\bibitem{abd_fru_integration_lopstr03}
S.~Abdennadher and T.~Fr{\"u}hwirth.
\newblock Integration and optimization of rule-based constraint solvers.
\newblock In M.~Bruynooghe, editor, {\em LOPSTR '03}, volume 3018 of {\em
  LNCS}, pages 198--213. Springer, 2004.

\bibitem{Abdennadher+99}
S.~Abdennadher, T.~Fr{\"u}hwirth, and H.~Meuss.
\newblock {C}onfluence and {S}emantics of {C}onstraint {S}implification
  {R}ules.
\newblock {\em Constraints}, 4(2):133--165, 1999.

\bibitem{AbMa00}
S.~Abdennadher and M.~Marte.
\newblock {U}niversity {C}ourse {T}imetabling {U}sing {C}onstraint {H}andling
  {R}ules.
\newblock In C.~Holzbaur and T.~Fr{\"u}hwirth, editors, {\em Special Issue on
  Constraint Handling Rules}, volume 14(4) of {\em Journal of Applied
  Artificial Intelligence}, pages 311--325. Taylor \& Francis, London, UK,
  2000.

\bibitem{abd_rigo_automatic_gen_of_solvers_tocl04}
S.~Abdennadher and C.~Rigotti.
\newblock Automatic generation of rule-based constraint solvers over finite
  domains.
\newblock {\em ACM TOCL}, 5(2):177--205, 2004.

\bibitem{generate}
S.~Abdennadher and C.~Rigotti.
\newblock Automatic generation of chr constraint solvers.
\newblock {\em Theory Pract. Log. Program.}, 5(4-5):403--418, 2005.

\bibitem{abd_saft_will_classroom_assignment_paclp00}
S.~Abdennadher, M.~Saft, and S.~Will.
\newblock Classroom assignment using constraint logic programming.
\newblock In {\em PACLP '00: Proc.\ 2nd Intl.\ Conf.\ and Exhibition on
  Practical Application of Constraint Technologies and Logic Programming}, Apr.
  2000.

\bibitem{AbdennadherSchuetz98}
S.~Abdennadher and H.~Sch{\"u}tz.
\newblock {CHR}v: {A} {F}lexible {Q}uery {L}anguage.
\newblock In {\em Third International Conference on Flexible Query Answering
  Systems}, volume 1495 of {\em LNCS}, pages 1--14. Springer, 1998.

\bibitem{aichernig_mutation_testing_2010}
B.~Aichernig.
\newblock A systematic introduction to mutation testing in unifying theories of
  programming.
\newblock In P.~Borba, A.~Cavalcanti, A.~Sampaio, and J.~Woodcook, editors,
  {\em Testing Techniques in Software Engineering}, volume 6153 of {\em LNCS},
  pages 243--287. Springer, 2010.

\bibitem{albert_clp_heap_iclp13}
E.~Albert, M.~J. {Garc{\'i}a de la Banda}, M.~G{\'o}mez-Zamalloa, J.~M. Rojas,
  and P.~J. Stuckey.
\newblock A {CLP} heap solver for test case generation.
\newblock volume 13(4--5) of {\em TPLP}, pages 721--735. Cambridge University
  Press, Aug. 2013.

\bibitem{alberti2008verifiable}
M.~Alberti, F.~Chesani, M.~Gavanelli, E.~Lamma, P.~Mello, and P.~Torroni.
\newblock Verifiable agent interaction in abductive logic programming: the
  sciff framework.
\newblock {\em ACM Transactions on Computational Logic (TOCL)}, 9(4):29, 2008.

\bibitem{AlDa+04}
M.~Alberti, D.~Daolio, P.~Torroni, M.~Gavanelli, E.~Lamma, and P.~Mello.
\newblock {S}pecification and {V}erification of {A}gent {I}nteraction
  {P}rotocols in a {L}ogic-{B}ased {S}ystem.
\newblock In {\em 2004 ACM {S}ymposium on {A}pplied {C}omputing}, pages 72--78.
  ACM, 2004.

\bibitem{alberti2013chr}
M.~Alberti, M.~Gavanelli, and E.~Lamma.
\newblock The {CHR}-based implementation of the sciff abductive system.
\newblock {\em Fundamenta Informaticae}, 124(4):365--381, 2013.

\bibitem{BaDo+04}
L.~Badea, D.~Tilivea, and A.~Hotaran.
\newblock {S}emantic {W}eb {R}easoning for {O}ntology-{B}ased {I}ntegration of
  {R}esources.
\newblock In {\em Second International Workshop on Principles and Practice of
  Semantic Web Reasoning}, volume 3208 of {\em LNCS}, pages 61--75. Springer,
  2004.

\bibitem{baryannis2014fluent}
G.~Baryannis and D.~Plexousakis.
\newblock Fluent calculus-based semantic web service composition and
  verification using wssl.
\newblock In {\em Service-Oriented Computing--ICSOC 2013 Workshops}, pages
  256--270. Springer, 2014.

\bibitem{betzPN}
H.~Betz.
\newblock {R}elating {C}oloured {P}etri {N}ets to {C}onstraint {H}andling
  {R}ules.
\newblock In {\em Fourth Workshop on Constraint Handling Rules}, pages 32--46,
  2007.

\bibitem{betz2014unified}
H.~Betz.
\newblock {\em A Unified Analytical Foundation for {Constraint Handling
  Rules}}.
\newblock BoD--Books on Demand, 2014.

\bibitem{betz2013linear}
H.~Betz and T.~Fr{\"u}hwirth.
\newblock Linear-logic based analysis of {Constraint Handling Rules} with
  disjunction.
\newblock {\em ACM Transactions on Computational Logic (TOCL)}, 14(1):1, 2013.

\bibitem{betz_raiser_fru_execution_model_iclp10}
H.~Betz, F.~Raiser, and T.~Fr{\"u}hwirth.
\newblock A complete and terminating execution model for {C}onstraint
  {H}andling {R}ules.
\newblock In Hermenegildo and Schaub \cite{piclp10}, pages 597--610.

\bibitem{bistarelli_fru_marte_rossi_soft_constraint_propagation_CI04}
S.~Bistarelli, T.~Fr{\"u}hwirth, M.~Marte, and F.~Rossi.
\newblock Soft constraint propagation and solving in {C}onstraint {H}andling
  {R}ules.
\newblock {\em Computational Intelligence: Special Issue on Preferences in AI
  and CP}, 20(2):287--307, May 2004.

\bibitem{bistarelli2010formal}
S.~Bistarelli, F.~Martinelli, and F.~Santini.
\newblock A formal framework for trust policy negotiation in autonomic systems:
  abduction with soft constraints.
\newblock {\em Autonomic and Trusted Computing}, pages 268--282, 2010.

\bibitem{DBLP:journals/cma/BistarelliMS12}
S.~Bistarelli, F.~Martinelli, and F.~Santini.
\newblock A semiring-based framework for the deduction/abduction reasoning in
  access control with weighted credentials.
\newblock {\em Computers {\&} Mathematics with Applications}, 64(4):447--462,
  2012.

\bibitem{buttner2012lightweight}
F.~B{\"u}ttner and J.~Cabot.
\newblock Lightweight string reasoning for {OCL}.
\newblock In {\em Modelling Foundations and Applications}, pages 244--258.
  Springer, 2012.

\bibitem{6274053}
R.~Chen, L.~Liao, and Z.~Fang.
\newblock Contracting of web services with {Constraint Handling Rules}.
\newblock In {\em Services (SERVICES), 2012 IEEE Eighth World Congress on},
  pages 211--218, 2012.

\bibitem{ChiCra+06}
W.-N. Chin, F.~Craciun, S.-C. Khoo, and C.~Popeea.
\newblock {A} {F}low-{B}ased {A}pproach for {V}ariant {P}arametric {T}ypes.
\newblock In {\em 21st annual ACM SIGPLAN Conference on Object-Oriented
  Programming Systems, Languages, and Applications}, pages 273--290. ACM, 2006.

\bibitem{christiansen2005chr}
H.~Christiansen.
\newblock Chr grammars.
\newblock {\em Theory and Practice of Logic Programming}, 5(4-5):467--501,
  2005.

\bibitem{christiansen2008implementing}
H.~Christiansen.
\newblock Implementing probabilistic abductive logic programming with
  {Constraint Handling Rules}.
\newblock In {\em Constraint Handling Rules}, pages 85--118. Springer, 2008.

\bibitem{christiansen_dahl_hyprolog_iclp05}
H.~Christiansen and V.~Dahl.
\newblock {HYPROLOG}: A new logic programming language with assumptions and
  abduction.
\newblock In Gabbrielli and Gupta \cite{piclp05}, pages 159--173.

\bibitem{christiansenconfluence}
H.~Christiansen and M.~H. Kirkeby.
\newblock Confluence modulo equivalence in {Constraint Handling Rules}.
\newblock  \cite{lopstr2014}.

\bibitem{lm-rocha2015}
F.~Cruz and R.~Rocha.
\newblock On compiling linear logic programs with comprehensions, aggregates
  and rule priorities.
\newblock In {\em Practical Aspects of Declarative Languages - 17th
  International Symposium, {PADL} 2015 Proceedings}, 2015.

\bibitem{dekoninck_la_meets_chr_tplp09}
L.~De~Koninck.
\newblock Logical {A}lgorithms meets {CHR}: A meta-complexity result for
  {C}onstraint {H}andling {R}ules with rule priorities.
\newblock {\em TPLP}, 9(2):165--212, Mar. 2009.

\bibitem{dekoninck_schr_demoen_search_chr06}
L.~De~Koninck, T.~Schrijvers, and B.~Demoen.
\newblock Search strategies in {CHR(Prolog)}.
\newblock In Schrijvers and Fr{\"u}hwirth \cite{pchr06}, pages 109--124.

\bibitem{dekoninck_schr_demoen_chrrp_techrep07}
L.~De~Koninck, T.~Schrijvers, and B.~Demoen.
\newblock Chr$^\mathrm{rp}$: {C}onstraint {H}andling {R}ules with rule
  priorties.
\newblock Technical Report CW 479, K.U.Leuven, Dept.\ Comp.\ Sc., Leuven,
  Belgium, Mar. 2007.

\bibitem{degrave_schr_vanhoof_test_inputs_merc_lopstr08}
F.~Degrave, T.~Schrijvers, and W.~Vanhoof.
\newblock Automatic generation of test inputs for mercury.
\newblock In M.~Hanus, editor, {\em LOPSTR '08, Revised Selected Papers},
  volume 5438 of {\em LNCS}. Springer, 2009.

\bibitem{giustu_gabbri_meo_multiple_heads_sofsem09}
C.~Di~Giusto, M.~Gabbrielli, and M.~C. Meo.
\newblock Expressiveness of multiple heads in {CHR}.
\newblock In {\em SOFSEM '09: Proc.\ 35th Conf.\ Current Trends in Theory and
  Practice of Comp.\ Science}, LNCS, pages 205--216. Springer, 2009.

\bibitem{pchr07}
K.~Djelloul, G.~J. Duck, and M.~Sulzmann, editors.
\newblock {\em CHR '07: Proc.\ 4th Workshop on Constraint Handling Rules},
  Sept. 2007.

\bibitem{DBLP:journals/corr/abs-1210-5307}
G.~J. Duck.
\newblock {SMCHR}: Satisfiability modulo {Constraint Handling Rules}.
\newblock {\em CoRR}, abs/1210.5307, 2012.

\bibitem{DBLP:journals/tplp/DuckHS14}
G.~J. Duck, R.~Haemmerl{\'{e}}, and M.~Sulzmann.
\newblock On termination, confluence and consistent {CHR}-based type inference.
\newblock {\em {TPLP}}, 14(4-5):619--632, 2014.

\bibitem{DBLP:conf/cp/DuckJK13}
G.~J. Duck, J.~Jaffar, and N.~C.~H. Koh.
\newblock Constraint-based program reasoning with heaps and separation.
\newblock In C.~Schulte, editor, {\em CP}, volume 8124 of {\em Lecture Notes in
  Computer Science}, pages 282--298. Springer, 2013.

\bibitem{duck_cadmium_iclp08}
G.~J. Duck, L.~D. Koninck, and P.~J. Stuckey.
\newblock Cadmium: An implementation of {ACD} term rewriting.
\newblock In Garc{\'i}a de~la Banda and Pontelli \cite{piclp08}, pages
  531--545.

\bibitem{duck_stuck_garc_holz_refined_op_sem_iclp04}
G.~J. Duck, P.~J. Stuckey, M.~{Garc{\'i}a de la Banda}, and C.~Holzbaur.
\newblock The refined operational semantics of {C}onstraint {H}andling {R}ules.
\newblock In B.~Demoen and V.~Lifschitz, editors, {\em ICLP '04}, volume 3132
  of {\em LNCS}, pages 90--104. Springer, Sept. 2004.

\bibitem{duck_stuck_sulz_observable_confluence_iclp07}
G.~J. Duck, P.~J. Stuckey, and M.~Sulzmann.
\newblock Observable confluence for {C}onstraint {H}andling {R}ules.
\newblock In V.~Dahl and I.~Niemel\"a, editors, {\em ICLP '07}, volume 4670 of
  {\em LNCS}, pages 224--239. Springer, Sept. 2007.

\bibitem{EsTo98}
M.~Escrig and F.~Toledo.
\newblock {\em {Q}ualitative {S}patial {R}easoning: {T}heory and {P}ractice}.
\newblock IOS Press, 1998.

\bibitem{pwlp06}
M.~Fink, H.~Tompits, and S.~Woltran, editors.
\newblock {\em WLP '06: Proc.\ 20th Workshop on Logic Programming}, T.U.Wien,
  Austria, INFSYS Research report 1843-06-02, Feb. 2006.

\bibitem{fru_chr_overview_jlp98}
T.~Fr{\"u}hwirth.
\newblock Theory and practice of {C}onstraint {H}andling {R}ules.
\newblock {\em J. Logic Programming, Special Issue on Constraint Logic
  Programming}, 37(1--3):95--138, 1998.

\bibitem{Fruehwirth00}
T.~Fr{\"u}hwirth.
\newblock {P}roving {T}ermination of {C}onstraint {S}olver {P}rograms.
\newblock In {\em Selected Papers from the Joint ERCIM/Compulog Net Workshop on
  New Trends in Contraints}, volume 1865 of {\em LNCS}, pages 298--317.
  Springer, 2000.

\bibitem{FruKR02}
T.~Fr{\"u}hwirth.
\newblock {A}s {T}ime {G}oes {B}y: {A}utomatic {C}omplexity {A}nalysis of
  {S}implification {R}ules.
\newblock In {\em Eighth International Conference on Principles of Knowledge
  Representation and Reasoning}, San Francisco, CA, USA, 2002. Morgan Kaufmann.

\bibitem{Fruehwirth2002}
T.~Fr{\"u}hwirth.
\newblock {A}s {T}ime {G}oes {B}y {II}: {M}ore {A}utomatic {C}omplexity
  {A}nalysis of {C}oncurrent {R}ule {P}rograms.
\newblock {\em ENTCS}, 59(3):185--206, 2002.

\bibitem{fru_parallel_union_find_iclp05}
T.~Fr{\"u}hwirth.
\newblock Parallelizing union-find in {C}onstraint {H}andling {R}ules using
  confluence.
\newblock In Gabbrielli and Gupta \cite{piclp05}, pages 113--127.

\bibitem{fru_specialization_lopstr04}
T.~Fr{\"u}hwirth.
\newblock Specialization of concurrent guarded multi-set transformation rules.
\newblock In S.~Etalle, editor, {\em LOPSTR '04}, volume 3573 of {\em LNCS},
  pages 133--148. Springer, 2005.

\bibitem{fru_general_union_find_csclp07}
T.~Fr{\"u}hwirth.
\newblock Quasi-linear-time algorithms by generalisation of union-find in
  {CHR}.
\newblock In {\em Recent Advances in Constraints --- CSCLP '07: 12th ERCIM
  Intl. Workshop on Constraint Solving and Constraint Logic Programming,
  Revised Selected Papers}, pages 91--118, Nov. 2008.

\bibitem{fru_welcome_lnai08}
T.~Fr{\"u}hwirth.
\newblock Welcome to {C}onstraint {H}andling {R}ules.
\newblock In Schrijvers and Fr{\"u}hwirth \cite{lnai08}, pages 1--15.

\bibitem{chr-book}
T.~Fr{\"u}hwirth.
\newblock {\em Constraint Handling Rules}.
\newblock Cambridge University Press, 2009.

\bibitem{fruhwirth2015constraint}
T.~Fr{\"u}hwirth.
\newblock Constraint handling rules -- what else?
\newblock In {\em Rule Technologies: Foundations, Tools, and Applications},
  pages 13--34. Springer International Publishing, 2015.

\bibitem{thom2015ppdp}
T.~Fr{\"u}hwirth.
\newblock A devil's advocate against termination of direct recursion.
\newblock In {\em 17th International Symposium on Principles and Practice of
  Declarative Programming, {PPDP} '15, Siena, Italy, 2015}. {ACM}, 2015.

\bibitem{fru_abd_munich_rent_advisor_tplp01}
T.~Fr\"{u}hwirth and S.~Abdennadher.
\newblock The {Munich} rent advisor: A success for logic programming on the
  internet.
\newblock {\em TPLP}, 1(3):303--319, 2001.

\bibitem{fru_abd_essentials_of_cp_book03}
T.~Fr{\"u}hwirth and S.~Abdennadher.
\newblock {\em Essentials of Constraint Programming}.
\newblock Springer, 2003.

\bibitem{fru_bri_base_wireless_ieee00}
T.~Fr{\"u}hwirth and P.~Brisset.
\newblock Placing base stations in wireless indoor communication networks.
\newblock {\em IEEE Intelligent Systems and Their Applications}, 15(1):49--53,
  2000.

\bibitem{fru02}
T.~Fr{\"u}hwirth, A.~di~Pierro, and H.~Wiklicky.
\newblock {P}robabilistic {C}onstraint {H}andling {R}ules.
\newblock In {\em 11th International Workshop on Functional and (Constraint)
  Logic Programming}, volume~76 of {\em ENTCS}, pages 115--130, 2002.

\bibitem{fru_holz_source2source_agp03}
T.~Fr{\"u}hwirth and C.~Holzbaur.
\newblock Source-to-source transformation for a class of expressive rules.
\newblock In F.~Buccafurri, editor, {\em AGP '03: Joint Conf.\ Declarative
  Programming APPIA-GULP-PRODE}, pages 386--397, Sept. 2003.

\bibitem{chr-thesis-book}
T.~Fr{\"u}hwirth and F.~Raiser, editors.
\newblock {\em Constraint Handling Rules: Compilation, Execution, and
  Analysis}.
\newblock BOD, 2011.

\bibitem{piclp05}
M.~Gabbrielli and G.~Gupta, editors.
\newblock {\em ICLP '05: Proc.\ 21st Intl.\ Conf.\ Logic Programming}, volume
  3668 of {\em LNCS}. Springer, Oct. 2005.

\bibitem{GabbrielliMM13}
M.~Gabbrielli, J.~Mauro, and M.~C. Meo.
\newblock The expressive power of {CHR} with priorities.
\newblock {\em Inf. Comput.}, 228:62--82, 2013.

\bibitem{gabbrielli_mauro_meo_sneyers_dedidability_iclp10}
M.~Gabbrielli, J.~Mauro, M.~C. Meo, and J.~Sneyers.
\newblock Decidability properties for fragments of {CHR}.
\newblock In Hermenegildo and Schaub \cite{piclp10}, pages 611--626.

\bibitem{gabbr_meo_compos_semantics_tocl09}
M.~Gabbrielli and M.~C. Meo.
\newblock A compositional semantics for {CHR}.
\newblock {\em ACM TOCL}, 10(2):1--36, Feb. 2009.

\bibitem{gabbrielli_unfolding_tplp13}
M.~Gabbrielli, M.~C. Meo, P.~Tacchella, and H.~Wiklicky.
\newblock Unfolding for {CHR} programs.
\newblock {\em Theory and Practice of Logic Programming}, pages 1--48, 2013.

\bibitem{gall2015lopstr}
D.~Gall and T.~Fr{\"u}hwirth.
\newblock A formal semantics for the cognitive architecture {ACT-R}.
\newblock  \cite{lopstr2014}.

\bibitem{gall2015ppdp}
D.~Gall and T.~Fr{\"u}hwirth.
\newblock A refined operational semantics for {ACT-R}.
\newblock In {\em 17th International Symposium on Principles and Practice of
  Declarative Programming, {PPDP} '15, Siena, Italy, 2015}. {ACM}, 2015.

\bibitem{piclp08}
M.~Garc{\'i}a de~la Banda and E.~Pontelli, editors.
\newblock {\em ICLP '08: Proc.\ 24rd Intl.\ Conf.\ Logic Programming}, volume
  5366 of {\em LNCS}. Springer, Dec. 2008.

\bibitem{gavanelli_alberti_lamma_abduction_optimization_ecai08}
M.~Gavanelli, M.~Alberti, and E.~Lamma.
\newblock Integrating abduction and constraint optimization in {C}onstraint
  {H}andling {R}ules.
\newblock In {\em ECAI 2008: 18th European Conf.\ on Artif.\ Intell.}, pages
  903--904. IOS press, July 2008.

\bibitem{geiselhart_et_al_mtseq_chr10}
F.~Geiselhart, F.~Raiser, J.~Sneyers, and T.~Fr{\"u}hwirth.
\newblock {MTSeq} -- multi-touch-enabled music generation and manipulation
  based on {CHR}.
\newblock In Van~Weert and De~Koninck \cite{pchr10}.

\bibitem{gerlich_atdg_chr10}
R.~Gerlich.
\newblock Generic and extensible {A}utomatic {T}est {D}ata {G}eneration for
  safety critical software with {CHR}.
\newblock In Van~Weert and De~Koninck \cite{pchr10}.

\bibitem{gerlich2014automatic}
R.~Gerlich.
\newblock Automatic test data generation and model checking with {CHR}.
\newblock {\em arXiv preprint arXiv:1406.2122}, 2014.

\bibitem{GevOs+01}
J.~Geurts, J.~V. Ossenbruggen, and L.~Hardman.
\newblock {A}pplication-{S}pecific {C}onstraints for {M}ultimedia
  {P}resentation {G}eneration.
\newblock In {\em 8th International Conference on Multimedia Modeling}, pages
  247--266, 2001.

\bibitem{gouraud_gotlieb_javacard_padl06}
S.-D. Gouraud and A.~Gotlieb.
\newblock Using {CHR}s to generate functional test cases for the {J}ava card
  virtual machine.
\newblock In P.~{Van Hentenryck}, editor, {\em PADL '06: Proc.\ 8th Intl.\
  Symp.\ Practical Aspects of Declarative Languages}, volume 3819 of {\em
  LNCS}, pages 1--15. Springer, Jan. 2006.

\bibitem{green2012logicblox}
T.~J. Green, M.~Aref, and G.~Karvounarakis.
\newblock Logicblox, platform and language: a tutorial.
\newblock In {\em Proceedings of the Second international conference on Datalog
  in Academia and Industry}, pages 1--8. Springer-Verlag, 2012.

\bibitem{haemmerle2011coinductive}
R.~Haemmerl{\'e}.
\newblock ({C}o-){I}nductive semantics for {C}onstraint {H}andling {R}ules.
\newblock volume 11(4--5) of {\em TPLP}, pages 593--609. Cambridge University
  Press, July 2011.

\bibitem{HaemmerlE:2012:DCC:2431176.2431194}
R.~Haemmerl{\'e}.
\newblock Diagrammatic confluence for {Constraint Handling Rules}.
\newblock {\em Theory Pract. Log. Program.}, 12(4-5):737--753, Sept. 2012.

\bibitem{haemmerle2011clp}
R.~Haemmerl{\'e}, P.~Lopez-Garcia, and M.~Hermenegildo.
\newblock {CLP} projection for constraint handling rules.
\newblock In M.~Hanus, editor, {\em PPDP '11}, pages 137--148. ACM Press, July
  2011.

\bibitem{piclp10}
M.~Hermenegildo and T.~Schaub, editors.
\newblock {\em ICLP '10: Proc.\ 26th Intl.\ Conf.\ Logic Programming}, volume
  10(4--6) of {\em TPLP}.
\newblock Cambridge University Press, July 2010.

\bibitem{holz_fru_prolog_chr_compiler_aai00}
C.~Holzbaur and T.~Fr{\"u}hwirth.
\newblock A {P}rolog {C}onstraint {H}andling {R}ules compiler and runtime
  system.
\newblock volume 14(4) of {\em Journal of Applied Artificial Intelligence},
  pages 369--388. Taylor \& Francis, Apr. 2000.

\bibitem{holz_garc_stuck_duck_opt_comp_chr_hal_tplp05}
C.~Holzbaur, M.~{Garc{\'i}a de la Banda}, P.~J. Stuckey, and G.~J. Duck.
\newblock Optimizing compilation of {C}onstraint {H}andling {R}ules in {HAL}.
\newblock volume 5(4--5) of {\em Theory and Practice of Logic Programming},
  pages 503--531. Cambridge University Press, July 2005.

\bibitem{LACHR}
L.~D. Koninck, T.~Schrijvers, and B.~Demoen.
\newblock {T}he {C}orrespondence {B}etween the {L}ogical {A}lgorithms
  {L}anguage and {CHR}.
\newblock In {\em 23rd International Conference on Logic Programming}, volume
  4670 of {\em LNCS}, pages 209--223. Springer, 2007.

\bibitem{lamsulz08}
E.~Lam and M.~Sulzmann.
\newblock {P}arallel {E}xecution of {M}ulti-{S}et {C}onstraint {R}ewrite
  {R}ules.
\newblock In {\em Tenth International ACM SIGPLAN Symposium on Principles and
  Practice of Declarative Programming}. ACM, 2008.

\bibitem{lam_sulzmann_conc_goal_based_tplp09}
E.~S. Lam and M.~Sulzmann.
\newblock Concurrent goal-based execution of {C}onstraint {H}andling {R}ules.
\newblock {\em TPLP}, 11:841--879, 2009.

\bibitem{lam2015}
E.~S.~L. Lam, I.~Cervesato, and N.~Fatima.
\newblock Comingle: Distributed logic programming for decentralized mobile
  ensembles.
\newblock In {\em Coordination Models and Languages - 17th {IFIP} {WG} 6.1
  International Conference, {COORDINATION} 2015, Grenoble, France, 2015}, pages
  51--66, 2015.

\bibitem{langbein_routing_robsail11}
J.~Langbein, R.~Stelzer, and T.~Fr{\"u}hwirth.
\newblock A rule-based approach to long-term routing for autonomous sailboats.
\newblock In {\em Robotic Sailing 2011, Part V}, pages 195--204, 2011.

\bibitem{martinez2012general}
E.~Mart{\i}nez-Mart{\i}n, M.~T. Escrig, and A.~P. del Pobil.
\newblock A general qualitative spatio-temporal model based on intervals.
\newblock {\em Journal of Universal Computer Science}, 18(10):1343--1378, 2012.

\bibitem{lopstr2014}
H.~S. Maurizio~Proietti, editor.
\newblock {\em Logic-Based Program Synthesis and Transformation, 24th
  International Symposium, LOPSTR 2014. Revised Selected Papers}, volume 8981
  of {\em {LNCS}}. Springer, 2015.

\bibitem{mauro2014constraint}
J.~Mauro.
\newblock {\em Constraints Meet Concurrency}.
\newblock Springer, 2014.

\bibitem{meister_preflow_push_wlp06}
M.~Meister.
\newblock Fine-grained parallel implementation of the preflow-push algorithm in
  {CHR}.
\newblock In Fink et~al. \cite{pwlp06}, pages 172--181.

\bibitem{meister2007reconstructing}
M.~Meister and T.~Fr{\"u}hwirth.
\newblock Reconstructing almost-linear tree equation solving algorithms in
  {CHR}.
\newblock In {\em Proceedings of CSCLP 2007: Annual ERCIM Workshop on
  Constraint Solving and Constraint Logic Programming}, page 123, 2007.

\bibitem{montali2010abductive}
M.~Montali, P.~Torroni, F.~Chesani, P.~Mello, M.~Alberti, and E.~Lamma.
\newblock Abductive logic programming as an effective technology for the static
  verification of declarative business processes.
\newblock {\em Fundamenta Informaticae}, 102(3):325--361, 2010.

\bibitem{pettorossi2014program}
A.~Pettorossi, F.~Fioravanti, M.~Proietti, and E.~De~Angelis.
\newblock Program verification using {Constraint Handling Rules} and array
  constraint generalizations.
\newblock In {\em VPT 2014. Second International Workshop on Verification and
  Program Transformation, July 17-18, 2014, Vienna, Austria}, volume~28, pages
  3--18. EasyChair, 2014.

\bibitem{pilozzi_auto_term_proofs_iclp09}
P.~Pilozzi.
\newblock Automating termination proofs for {CHR}.
\newblock In P.~M. Hill and D.~S. Warren, editors, {\em ICLP '09}, volume 5649
  of {\em LNCS}, pages 504--508. Springer, July 2009.

\bibitem{pilozzi_improved_termination_lopstr11}
P.~Pilozzi and D.~De~Schreye.
\newblock Improved termination analysis of {CHR} using self-sustainability
  analysis.
\newblock In G.~Vidal, editor, {\em LOPSTR '11, Revised Selected Papers}, LNCS,
  2011.

\bibitem{pretschner_et_al_model_based_testing_jss04}
A.~Pretschner, H.~L{\"o}tzbeyer, and J.~Philipps.
\newblock Model based testing in incremental system development.
\newblock {\em Journal of Systems and Software}, 70(3):315--329, 2004.

\bibitem{raiser08}
F.~Raiser.
\newblock {G}raph {T}ransformation {S}ystems in {CHR}.
\newblock In {\em 23rd International Conference on Logic Programming}, volume
  4670 of {\em LNCS}, pages 240--254. Springer, 2007.

\bibitem{raiser_globalconstraintautomata_cp08}
F.~Raiser.
\newblock Semi-automatic generation of {CHR} solvers from global constraint
  automata.
\newblock In P.~J. Stuckey, editor, {\em CP' 08: Proc.\ 14th Intl.\ Conf.\
  Princ.\ Pract.\ Constraint Programming}, volume 5202 of {\em LNCS}, pages
  588--592. Springer, Sept. 2008.

\bibitem{raiser_betz_fru_equivalence_revisited_chr09}
F.~Raiser, H.~Betz, and T.~Fr{\"u}hwirth.
\newblock Equivalence of {CHR} states revisited.
\newblock In F.~Raiser and J.~Sneyers, editors, {\em CHR '09}, pages 33--48.
  K.U.Leuven, Dept.\ Comp.\ Sc., Technical report CW 555, July 2009.

\bibitem{raiser_fru_towardstermrewriting_chr08}
F.~Raiser and T.~Fr{\"u}hwirth.
\newblock Towards term rewriting systems in {C}onstraint {H}andling {R}ules.
\newblock In T.~Schrijvers, F.~Raiser, and T.~Fr{\"u}hwirth, editors, {\em CHR
  '08}, pages 19--34. RISC Report Series 08-10, University of Linz, Austria,
  2008.

\bibitem{raiser_fru_parallel_wlp10}
F.~Raiser and T.~Fr{\"u}hwirth.
\newblock Exhaustive parallel rewriting with multiple removals.
\newblock In S.~Abdennadher, editor, {\em WLP '10}, Sept. 2010.

\bibitem{raiser_tacchella_confluence_non_terminating_chr07}
F.~Raiser and P.~Tacchella.
\newblock On confluence of non-terminating {CHR} programs.
\newblock In Djelloul et~al. \cite{pchr07}, pages 63--76.

\bibitem{salomie2010web}
I.~Salomie, V.~Chifu, I.~Harsa, and M.~Gherga.
\newblock Web service composition using fluent calculus.
\newblock {\em International Journal of Metadata, Semantics and Ontologies},
  5(3):238--250, 2010.

\bibitem{sarnastarosta_ramakrishnan_chrd_padl07}
B.~Sarna-Starosta and C.~Ramakrishnan.
\newblock Compiling {C}onstraint {H}andling {R}ules for efficient tabled
  evaluation.
\newblock In M.~Hanus, editor, {\em PADL '07: Proc.\ 9th Intl.\ Symp.\
  Practical Aspects of Declarative Languages}, volume 4354 of {\em LNCS}, pages
  170--184. Springer, Jan. 2007.

\bibitem{ss_stirewalt_dillon_checking_deadlock_ijseke07}
B.~Sarna-Starosta, R.~E.~K. Stirewalt, and L.~K. Dillon.
\newblock A model-based design-for-verification approach to checking for
  deadlock in multi-threaded applications.
\newblock {\em Intl.\ Journal of Softw.\ Engin.\ and Knowl.\ Engin.},
  17(2):207--230, 2007.

\bibitem{Fluxpl07}
S.~Schiffel and M.~Thielscher.
\newblock {F}luxplayer: {A} {S}uccessful {G}eneral {G}ame {P}layer.
\newblock In {\em 22nd Conference on Artificial Intelligence}, pages
  1191--1196. AAAI Press, 2007.

\bibitem{SchriBru06}
T.~Schrijvers and M.~Bruynooghe.
\newblock {P}olymorphic {A}lgebraic {D}ata {T}ype {R}econstruction.
\newblock In {\em Eighth ACM SIGPLAN International Conference on Principles and
  Practice of Declarative Programming}, pages 85--96. ACM, 2006.

\bibitem{pchr06}
T.~Schrijvers and T.~Fr{\"u}hwirth, editors.
\newblock {\em CHR '06: Proc.\ 3rd Workshop on Constraint Handling Rules}.
  K.U.Leuven, Dept.\ Comp.\ Sc., Technical report CW 452, July 2006.

\bibitem{schr_fru_opt_union_find_tplp06}
T.~Schrijvers and T.~Fr\"{u}hwirth.
\newblock Optimal union-find in {C}onstraint {H}andling {R}ules.
\newblock {\em TPLP}, 6(1--2):213--224, 2006.

\bibitem{lnai08}
T.~Schrijvers and T.~Fr{\"u}hwirth, editors.
\newblock {\em {C}onstraint {H}andling {R}ules --- Current Research Topics},
  volume 5388 of {\em LNAI}.
\newblock Springer, Dec. 2008.

\bibitem{schr_sulz_transactions_iclp08}
T.~Schrijvers and M.~Sulzmann.
\newblock Transactions in {C}onstraint {H}andling {R}ules.
\newblock In Garc{\'i}a de~la Banda and Pontelli \cite{piclp08}, pages
  516--530.

\bibitem{SeBa+02}
C.~Seitz, B.~Bauer, and M.~Berger.
\newblock {M}ulti {A}gent {S}ystems {U}sing {C}onstraint {H}andling {R}ules for
  {P}roblem {S}olving.
\newblock In {\em International Conference on Artificial Intelligence}, pages
  295--301. CSREA Press, 2002.

\bibitem{sneyers_subclass_iclp08}
J.~Sneyers.
\newblock {T}uring-complete subclasses of {CHR}.
\newblock In Garc{\'i}a de~la Banda and Pontelli \cite{piclp08}, pages
  759--763.

\bibitem{apopcaleaps}
J.~Sneyers and D.~{De Schreye}.
\newblock {APOPCALEAPS}: Automatic music generation with {CHRiSM}.
\newblock In G.~Danoy et~al., editors, {\em 22nd Benelux Conference on
  Artificial Intelligence (BNAIC 2010)}, Luxembourg, October 2010.

\bibitem{sneyers2013probabilistic}
J.~Sneyers, D.~De~Schreye, and T.~Fr{\"u}hwirth.
\newblock Probabilistic legal reasoning in {CHRiSM}.
\newblock {\em Theory and Practice of Logic Programming}, 13(4-5):769--781,
  2013.

\bibitem{sneyers2010chr}
J.~Sneyers, W.~Meert, J.~Vennekens, Y.~Kameya, and T.~Sato.
\newblock Chr ({PRISM})-based probabilistic logic learning.
\newblock {\em Theory and Practice of Logic Programming}, 10(4-6):433--447,
  2010.

\bibitem{sney_schr_demoen_dijkstra_chr_wlp06}
J.~Sneyers, T.~Schrijvers, and B.~Demoen.
\newblock Dijkstra's algorithm with {Fibonacci} heaps: An executable
  description in {CHR}.
\newblock In Fink et~al. \cite{pwlp06}, pages 182--191.

\bibitem{sney_schr_demoen_chr_complexity_toplas09}
J.~Sneyers, T.~Schrijvers, and B.~Demoen.
\newblock The computational power and complexity of {C}onstraint {H}andling
  {R}ules.
\newblock {\em ACM TOPLAS}, 31(2), Feb. 2009.

\bibitem{sney_vanweert_demoen_aggregates_chr07}
J.~Sneyers, P.~Van~Weert, and T.~Schrijvers.
\newblock Aggregates for {C}onstraint {H}andling {R}ules.
\newblock In Djelloul et~al. \cite{pchr07}, pages 91--105.

\bibitem{chr_survey_tplp10}
J.~Sneyers, P.~Van~Weert, T.~Schrijvers, and L.~De~Koninck.
\newblock As time goes by: {C}onstraint {H}andling {R}ules -- {A} survey of
  {CHR} research between 1998 and 2007.
\newblock {\em TPLP}, 10(1):1--47, 2010.

\bibitem{sobhi_abd_betz_intentionally_def_constraints_lnai08}
I.~Sobhi, S.~Abdennadher, and H.~Betz.
\newblock Constructing rule-based solvers for intentionally-defined
  constraints.
\newblock In Schrijvers and Fr{\"u}hwirth \cite{lnai08}, pages 70--84.

\bibitem{StuSu05}
P.~J. Stuckey and M.~Sulzmann.
\newblock {A} {T}heory of {O}verloading.
\newblock {\em ACM Transactions on Programming Languages and Systems},
  27(6):1216--1269, 2005.

\bibitem{sulz_duck_peyton_stuck_func_dep_via_chr_fp07}
M.~Sulzmann, G.~J. Duck, S.~Peyton-Jones, and P.~J. Stuckey.
\newblock Understanding functional dependencies via {C}onstraint {H}andling
  {R}ules.
\newblock {\em J. Functional Prog.}, 17(1):83--129, 2007.

\bibitem{Flux05}
M.~Thielscher.
\newblock {FLUX}: {A} {L}ogic {P}rogramming {M}ethod for {R}easoning {A}gents.
\newblock {\em Theory and Practice of Logic Programming}, 5:533--565, 2005.

\bibitem{thielscher2006reasoning}
M.~Thielscher.
\newblock {\em Reasoning robots: the art and science of programming robotic
  agents}, volume~33.
\newblock Springer Science \& Business Media, 2006.

\bibitem{triossi2012compiling}
A.~Triossi, S.~Orlando, A.~Raffaet{\`a}, and T.~Fr{\"u}hwirth.
\newblock Compiling chr to parallel hardware.
\newblock In {\em Proceedings of the 14th symposium on Principles and practice
  of declarative programming}, pages 173--184. ACM, 2012.

\bibitem{vanweert_jchr_compilation_techrep08}
P.~Van~Weert.
\newblock Compiling {C}onstraint {H}andling {R}ules to {J}ava: A
  reconstruction.
\newblock Technical Report CW 521, K.U.Leuven, Dept.\ Comp.\ Sc., Leuven,
  Belgium, Aug. 2008.

\bibitem{vanweert_lazy_evaluation_tkde10}
P.~Van~Weert.
\newblock Efficient lazy evaluation of rule-based programs.
\newblock {\em IEEE Transactions on Knowledge and Data Engineering},
  22(11):1521--1534, Nov. 2010.

\bibitem{pchr10}
P.~Van~Weert and L.~De~Koninck, editors.
\newblock {\em CHR '10: Proc.\ 7th Workshop on Constraint Handling Rules}.
  K.U.Leuven, Dept.\ Comp.\ Sc., Technical report CW 588, July 2010.

\bibitem{vanweert_sney_schr_demoen_negation_chr06}
P.~Van~Weert, J.~Sneyers, T.~Schrijvers, and B.~Demoen.
\newblock Extending {CHR} with negation as absence.
\newblock In Schrijvers and Fr{\"u}hwirth \cite{pchr06}, pages 125--140.

\bibitem{vanweert_wuille_et_al_chr_imperative_lnai08}
P.~Van~Weert, P.~Wuille, T.~Schrijvers, and B.~Demoen.
\newblock {CHR} for imperative host languages.
\newblock In Schrijvers and Fr{\"u}hwirth \cite{lnai08}, pages 161--212.

\bibitem{wolf2005intelligent}
A.~Wolf.
\newblock Intelligent search strategies based on adaptive {Constraint Handling
  Rules}.
\newblock {\em Theory and Practice of Logic Programming}, 5(4-5):567--594,
  2005.

\bibitem{wolf_robin_vitorino_chrv_semantics_lnai08}
A.~Wolf, J.~Robin, and J.~Vitorino.
\newblock Adaptive {CHR} meets {CHR$^\vee$}: An extended refined operational
  semantics for {CHR$^\vee$} based on justifications.
\newblock In Schrijvers and Fr{\"u}hwirth \cite{lnai08}, pages 48--69.

\bibitem{wuille_schr_demoen_cchr_chr07}
P.~Wuille, T.~Schrijvers, and B.~Demoen.
\newblock {CCHR}: the fastest {CHR} implementation, in {C}.
\newblock In Djelloul et~al. \cite{pchr07}, pages 123--137.

\bibitem{zaki_parallel_gpu_chr12}
A.~Zaki, T.~Fr{\"u}hwirth, and I.~Geller.
\newblock Parallel execution of {C}onstraint {H}andling {R}ules on a
  {G}raphical {P}rocessing {U}nit.
\newblock In J.~Sneyers and T.~Fr{\"u}hwirth, editors, {\em CHR '12}, pages
  82--90. K.U.Leuven, Dept.\ Comp.\ Sc., Technical report CW 624, Sept. 2012.

\bibitem{zhu_madnick_siegel_IJEB08}
H.~Zhu, S.~E. Madnick, and M.~D. Siegel.
\newblock Enabling global price comparison through semantic integration of web
  data.
\newblock {\em IJEB}, 6(4):319--341, 2008.

\end{thebibliography}

\end{document}